\documentclass[apjl]{emulateapj}
\usepackage{graphicx}

\newcommand{\e}[1]{10$^{#1}$}
\newcommand{\ee}[1]{$\times$10$^{#1}$}

\newcommand{\phos}{~photons\,\,cm$^{-2}$\,s$^{-1}$}

\newcommand{\msun}{M$_{\odot}$}

\shorttitle{BAT observations of Tycho's SNR}
\shortauthors{E. Troja et al.}

\begin{document}
\title{
Swift/BAT detection of hard X-rays from Tycho's Supernova Remnant: \mbox{Evidence for Titanium-44} }

\author{E.~Troja\altaffilmark{1,2}, A. Segreto\altaffilmark{3}, V.~La Parola\altaffilmark{3},
D. Hartmann\altaffilmark{4}, W. Baumgartner\altaffilmark{1}, C. Markwardt\altaffilmark{1},
S. Barthelmy\altaffilmark{1}, G. Cusumano\altaffilmark{3}, and N. Gehrels\altaffilmark{1}
}
	 
\altaffiltext{1}{NASA, Goddard Space Flight Center, Greenbelt, MD}	
\altaffiltext{2}{Department of Physics and Astronomy, University of Maryland, College Park, MD}	
\altaffiltext{3}{INAF - IASF Palermo, via Ugo La Malfa, Italy}	
\altaffiltext{4}{Department of Physics and Astronomy, Clemson University, Clemson, SC}

\begin{abstract}
We report {\it Swift}/BAT survey observations of the Tycho's supernova remnant, 
performed over a period of 104 months since the mission's launch. 
The remnant is detected with high significance ($>$10\,$\sigma$) below 50~keV. 
We detect significant hard X-ray emission in the 60-85~keV band, 
above the continuum level predicted by a simple synchrotron model.
The location of the observed excess is consistent with line emission from radioactive Titanium-44,
so far reported only for Type II supernova explosions.
We discuss the implications of these results in the context of the galactic supernova rate, 
and nucleosynthesis in Type Ia supernova. 

\end{abstract}


\section{Introduction}\label{sec:intro}

Radioactive elements produced during a supernova (SN) explosion
carry unique information about the explosive nucleosynthesis process,
the stellar progenitor, and the explosion mechanism itself.
An element of great astrophysical interest is Titanium-44, which,
with a half-life of $\sim$59~yr \citep{ahmad06}, can give rise to observable 
features in the spectra of young supernova remnants (SNRs). 
The $^{44}$Ti$\rightarrow$$^{44}$Sc$\rightarrow$$^{44}$Ca decay chain 
produces three de-excitation lines of 
roughly equal branching ratios, at 1157~keV (from $^{44}$Ca), 
and at 78.4~keV, and 67.9~keV (from $^{44}$Sc) .
Observations of young SNRs in the gamma-ray and 
hard X-ray bands therefore represent a promising window into the latest stages 
of stellar evolution \citep{clayton69,leising01,vink2012}.

The first direct evidence of $^{44}$Ti was found by the
{\it Compton Gamma Ray Observatory} \citep{cgro94} in the SNR Cassiopea~A.
The daughter $^{44}$Sc emission lines were later detected in the hard X-ray band
by {\it Beppo-SAX} \citep{vink2001}, and {\it INTEGRAL} \citep{renaud06}.
The observed flux of $\approx$2\ee{-5}\phos~in each line implies a 
$^{44}$Ti mass of $\approx$2\ee{-4}\,\msun,
consistent with a core-collapse origin \citep{krause08,chieffi13}.
The spatial distribution of the Scandium lines,
as recently imaged by {\it NuSTAR} \citep{harrison13}, provides strong evidence 
for an asymmetric explosion,
likely caused by low-mode convection \citep{nustar14}.

For SNe Type Ia, sub-Chandrasekhar mass models \citep{ww94} 
can synthesize large amounts of radioactive of $^{44}$Ti, 
exceeding \e{-3}\,\msun in some scenarios \citep{fink10,wk11}.
The $^{44}$Ti yield is probably less abundant in normal Type~Ia~explosions, 
ranging from $\sim$\e{-6}\,\msun~for a centrally-ignited pure-deflagration
to $\lesssim$6\ee{-5}\,\msun~for an off-center delayed detonation \citep{iwamoto99,maeda10}.
Indeed, previous attempts to detect radioactive lines
in SN Ia remnants were unsuccessful \citep{dupraz97, renaud2}.
The historical Tycho remnant, originated in 1572AD from a Type Ia SN \citep{baade45, krausetycho}, 
has been a primary target for searching nuclear emission lines. 
A 1.5~Ms long observation by {\it INTEGRAL} detected the X-ray continuum 
only up to energies of $\sim$50~keV, placing a 3$\sigma$ upper limit of 
1.5\ee{-5}\phos~on the flux of the $^{44}$Sc lines \citep{renaud2,wangli14}.

In this Letter we report first evidence of significant emission 
above 50~keV from the Tycho SNR, 
detected by the Burst Alert Telescope \citep[BAT;][]{scott05} 
on-board {\it Swift} \citep{neil04}. 
The signal is detected at a $\sim$4 sigma confidence level
in the 60-85~keV energy band, consistent with the location of the 
Scandium lines.
Our observations and data analysis are described in Section~\ref{sec:data}. 
Our results are discussed in Section~\ref{sec:result}.
Unless otherwise stated, the quoted uncertainties represent the 90\% confidence
interval for one interesting parameter, corresponding to $\Delta \chi^2$=2.706 \citep{lampton76}. 

\begin{figure*}[!t]
\includegraphics[scale=1]{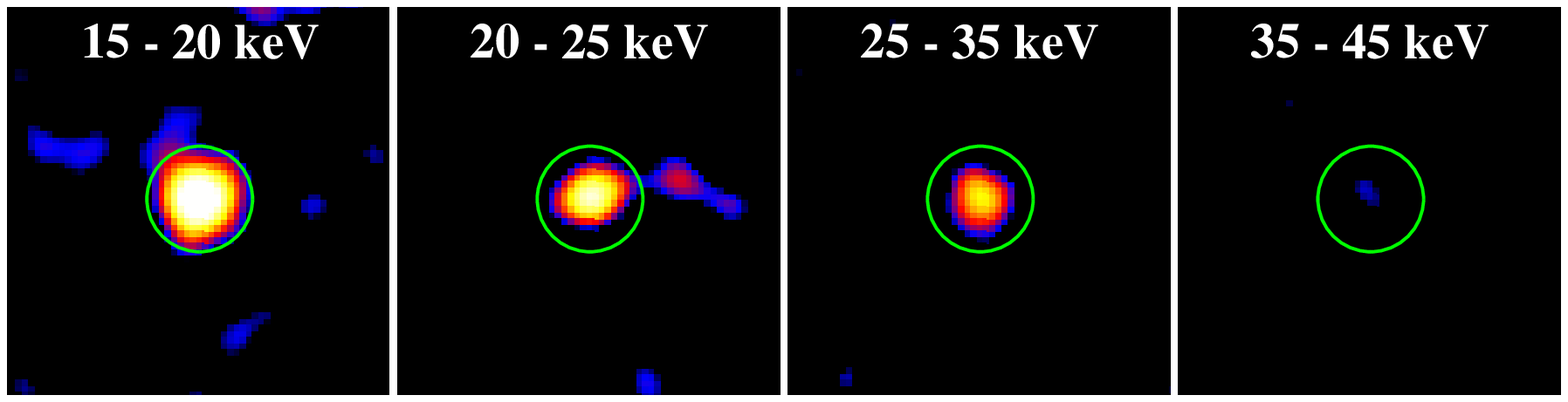}
\includegraphics[scale=0.96]{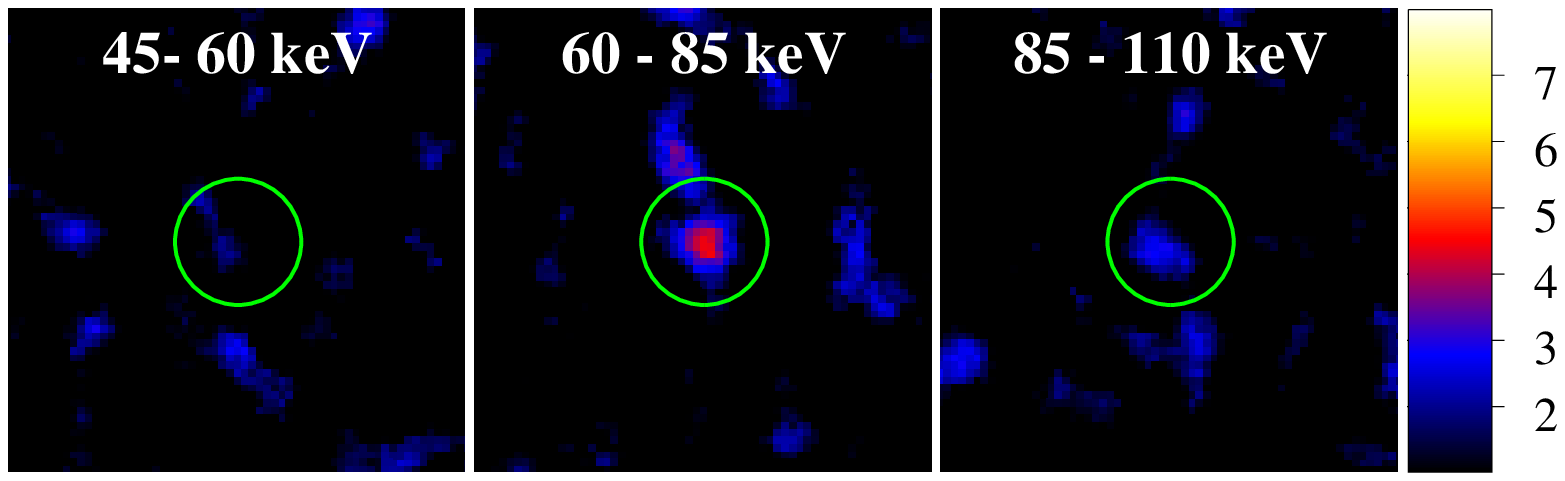}
\caption{{\it Swift}/BAT image of Tycho's SNR in different energy ranges. 
The SNR position is indicated by the green circle. 
The color scale is proportional to the detection significance 
in units of standard deviations.}
\end{figure*}

\section{Data analysis}\label{sec:data}

\subsection{Observations and Data Reduction}

Since its launch in November 2004, {\it Swift} is performing an 
all-sky survey with the BAT, covering the 14-195~keV energy range. 
Thanks to the BAT's wide field of view (1.4~sr, half-coded), 
its large detector area (5432 cm$^2$),
and  the random {\it Swift} pointing strategy, 
the BAT Survey resulted in the most sensitive and uniform coverage
of the hard X-ray sky \citep{bat22,palermo10,bat70}. 

We analyzed the data from November 2004 until July 2013, spanning a period of 104
months since the mission's launch.
The BAT survey raw data were retrieved from the {\it Swift} public
archive\footnote{http://heasarc.gsfc.nasa.gov/cgi-bin/W3Browse/swift.pl}
where they are stored in the form of detector plane histograms:
three-dimensional arrays (two spatial dimensions, one spectral
dimension) that collect count-rate data in 5-min time bins for
80 energy channels. These data were processed with a dedicated software 
\citep{segreto10}
that computes all-sky maps in 8 energy bands between 15 and 150 keV,
performs source detection on these maps, and for each detected source
produces standard products such as background subtracted light curves and spectra.

The Tycho's SNR is detected in the 15--150 keV all-sky map
with a signal to noise ratio of 15.6 standard deviations.
The source was inside the BAT field of view for approximately 43 Ms, 
corresponding to a total on-axis equivalent exposure of 19.6~Ms.
An image of the SNR in different energy channels is presented in Figure~1. 
The source is detected with high significance in the low energy channels
($<$35~keV), and only marginally in the 35-45 keV energy range. 
The detection significance in the 60-85~keV bin has been growing with time, 
from 3.4 $\sigma$ (using the first 25 months of survey data), 
to 3.9 $\sigma$ (50 months), and to 4.7 $\sigma$ (75 months). 

An independent analysis, based on the products of the standard GSFC pipeline \citep{bat22,bat70},
finds consistent results. The GSFC processing \citep{bat70} covers the period from November 2004 through
August 2013, for a total on-axis exposure of 19.6~Ms. Data are rebinned into eight survey energy bands between 14~keV and 195~keV.
The SNR is detected with high-significance below 35~keV, and with lower significance in the 
50-75~keV (2.4~$\sigma$), and 75-100~keV (3.5~$\sigma$) energy bands. 

As the detected signal above 50~keV is faint, 
background modeling and subtraction is of critical importance in its analysis.
The analysis technique for BAT coded mask data inherently subtracts the background,
as discussed in more detail in \citet{segreto10} and \citet{bat70}.
Imperfections in subtraction of non-imaged background are apparent as 
pattern noise in the resulting all-sky images from which the SNR flux values are derived.  The significance of the flux properly 
accounts for pattern noise, in addition to Poisson noise.

\subsection{Spectral analysis}\label{sec:pha}

The continuum shape is a critical element
for determing the nature of the emission above 50~keV.
The most likely origin of the hard X-ray continuum in the Tycho's SNR 
is synchrotron radiation from a population of shock-accelerated
electrons.
Both {\it Suzaku} \citep{suzaku} and {\it Chandra} \citep{hwang02,chandra}
observations are consistent with a simple power-law of photon
index $\Gamma$$\approx$2.8. However, the SNR's broadband spectrum
deviates from a simple power-law, showing a gradual turn-over 
at energies $E$\,$\approx$1 keV.
In this case, a simple power-law model might overestimate the contribution
of the continuum above 60~keV.
We therefore modelled the continuum shape by using a power-law function
as well as a synchrotron cutoff model  (model \texttt{srcut}; \citealt{rk99}) 
with low-energy slope $\alpha$=0.65, and a 1.4~GHz flux density of 40.5~Jy \citep{kothes06}.

\begin{figure}[!t]
\centering
\includegraphics[scale=0.43]{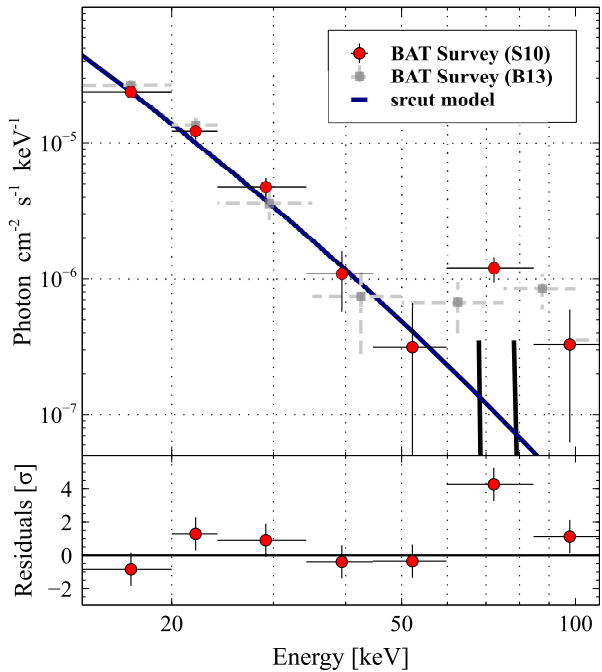}
\caption{BAT spectrum of the Tycho's SNR. We show the survey data processed
with the \citet{segreto10} pipeline (S10), used for the spectral analysis, 
and, for comparison, the survey data processed with the GSFC pipeline (B13).
The best fit model for the hard X-ray continuum is a syncrotron model (srcut function), 
shown as a solid line. The two vertical lines mark the energies
of the $^{44}$Sc lines.
The bottom panel shows the fit residuals in units of sigma.
}
\end{figure}

Spectral fits were performed with XSPEC v.12.8.1 by minimizing the 
$\chi^2$ statistics. All the fits were carried out in the 15-110~keV energy band.
The power-law fit yields a photon index $\Gamma$=3.0$\pm$0.3,
 consistent with previous {\it Chandra} and {\it Suzaku} results,
and a $\chi^2$=18 for 5 degrees of freedom (d.o.f.).
The srcut fit yields a $\chi^2$=20 for 6 d.o.f. and a 
break frequency $\nu_{b}$=(2.00$\pm$0.12)\ee{17}~Hz,
 in agreement with the values quoted by \citet{suzaku} and \citet{chandra}.
The poor fits are mainly due to the signal detected in the 60-85~keV bin,
which lies 4~$\sigma$ above the predicted continuum decay. 
We modelled the observed excess with two gaussian profiles
at fixed centroids of 68~keV and 78 keV, zero width,
and relative flux ratio F$_{68}$=0.93 F$_{78}$. 
The resulting fits are acceptable, with the srcut model 
($\chi^2$=6.2 for 5 d.o.f.) being statistically preferred over
the simple power-law ($\chi^2$=7.4 for 4 d.o.f.).
The SNR spectrum and the best fit model (without the contribution of the $^{44}$Sc lines)
are reported in Figure~2.
The derived fluxes for the two lines are F$_{78}$=(1.4$\pm$0.6)\ee{-5}\,\phos,
and F$_{68}$=(1.3$\pm$0.6)\ee{-5}\,\phos.
We further investigated the dependence of the lines flux
on the underlying continuum shape. 
Figure 3 shows the confidence level contours of the 78 keV line flux
as a function of the break frequency.  The resulting 3\,$\sigma$ error range is 
2.1\ee{-6} $<$F$_{78}$$<$2.6\ee{-5}\,\phos.

A different possibility is that the observed excess is due to an additional
continuum component, such as non-thermal bremsstrahlung emission.
We therefore added to our continuum model a power-law component. 
The resulting fit with a synchrotron+power-law model is poor ($\chi^2$=12 for 4 d.o.f.).
In fact, the power-law flux in the 60-85~keV bin is constrained to 
$\lesssim$0.8\ee{-6}\,\phos~by the non-detections in the adjacent 45-60~keV, 
and 85-110~keV bins, and cannot account for all the observed emission.

In order to check whether the observed excess in the 65-80 keV band could be 
an artifact we performed the following tests:
1) after excluding the point sources, we checked that the statistical distribution of the 
skymap pixels significances is a gaussian distribution with mean 0 and standard deviation 1;
2) we checked that spectra extracted at random source-free positions did not present 
any feature, and were consistent with zero;
3) we checked that the detected sources located near the Tycho's SNR did not present the same 
65-80 keV excess in their spectra, nor any other peculiar feature.
In particular, this last point allows us to exclude that the observed feature is a 
systematic effect related to the response matrix.

\begin{figure}[!t]
\centering
\includegraphics[scale=0.37,angle=270, trim=1.5cm 0.5cm 0cm 0cm, clip=true]{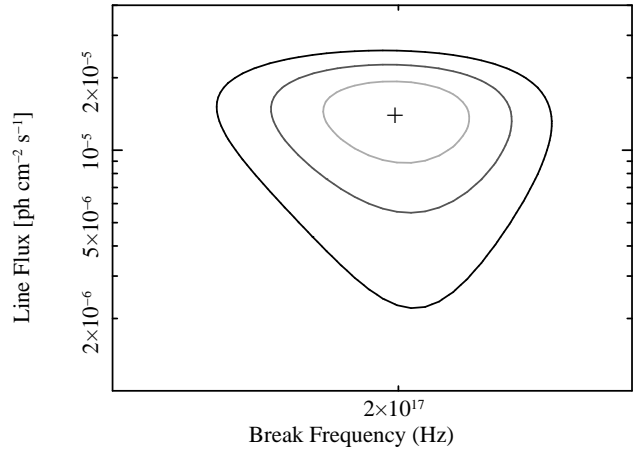}
\caption{Confidence contours for the line flux $F_{78}$ and the
synchrotron frequency $\nu_b$. The best fit value is marked by the cross
symbol. The confidence levels are 1, 2, and 3\,$\sigma$ for 
two parameters of interest, corresponding to $\Delta \chi^2$=4.61, 6.17, and 11.8 \citep{lampton76}.}
\end{figure}

\subsubsection{Consistency check: the SNR Cas~A}\label{casa}

The same data reduction and analysis was applied to the 
BAT spectrum of the SNR Cas~A, which is a known source of $^{44}$Ti.
The SNR was within the BAT FoV for 44~Ms, corresponding to
an on-axis equivalent exposure of 19.9~Ms.
The signal in the 60-85~keV range is detected with high significance
($>$13~$\sigma$).

Also in this case, as expected, the fit with a simple power-law function 
is not acceptable ($\chi^2$=68 for 5 d.o.f.), as it significantly underestimates
the flux in the 60-85~kev bin.
We therefore modelled the spectrum with a power-law function for the continuum,
and two narrow gaussian features at the position of the $^{44}$Sc lines.
This model provides a good description of the dataset ($\chi^2$=5.4 for 4 d.o.f.),
with best fit values $\Gamma$=3.32$\pm$0.07 for the power-law photon index, 
$A$=(1.5$\pm$0.3)\ee{-5}\,\phos at 1~keV for the power-law normalization,
F$_{78}$=(2.7$\pm$0.6)\ee{-5}\,\phos, and F$_{68}$=(2.5$\pm$0.6)\ee{-5}\,\phos\ 
for the fluxes of the two lines.
The derived fluxes are in excellent agreement with previous measurements
from Beppo-SAX \citep{vink2001}, and INTEGRAL \citep{renaud06}, and 
slightly higher than the value measured by {\it NuSTAR} \citep{nustar14}.
The observed X-ray emission from $^{44}$Sc is indeed expected to decay with time.
As our data span a period of $\sim$9 years prior to the {\it NuSTAR} observations, 
the expected decrease in the line flux is $<$10\%, and can partially account
for the observed difference. Another possibility is that a small fraction of the 
observed flux comes from a spatially diffuse component of emission, which was not
imaged by {\it NuSTAR}.

\newpage
\section{Results}~\label{sec:result}
The most notable feature of the BAT observations is the detection of
emission in the 60--85~keV range. 
The SNR spectrum is well described by a simple synchrotron model, which also 
provides the best description of the broadband (from radio to hard X-rays) 
emission \citep[e.g.][]{slane14}. However, this model fails to reproduce the 
observed signal above 60~keV.
The BAT non-detection above 85~keV constrains the contribution 
of any additional continuum component to be negligible. 
This is in agreement with the SNR properties (low gas density, and high magnetic field) 
inferred by other studies, which also imply a weak non-thermal bremsstrahlung. 

Although the broad spectral bins do not allow us to clearly resolve the two emission lines, 
the energy range of the observed excess is remarkably consistent with the location of the 
$^{44}$Sc nuclear lines. 
By attributing the observed signal above 60 keV to the $^{44}$Sc lines, we 
can estimate the $^{44}$Ti yield as \citep[e.g.][]{grebenev12}:
\begin{equation}
\frac{M({\rm Ti})}{M_{\odot}} = 1.41 \times 10^{-4} 
\left[ f_X \left( \frac{d}{\rm 1~kpc} \right)^2 
\tau~ {\rm exp}\left(\frac{T}{\tau}\right)~W^{-1} \right],
\end{equation}
where $f_X$ is the observed line flux in units of \phos, 
$d$ the SNR distance,
$T$ its age, and $\tau$=85.3$\pm$0.4~yr the $^{44}$Ti lifetime \citep{ahmad06}.
The emission efficiencies for the two lines are $W_{68}$=0.877, and  $W_{78}$=0.947, 
respectively. 

As shown in Equation~1, the estimated mass sensitively depends on the 
SNR distance. Various measurements constrain $d$ in the range 
between 1.5 and 5 kpc, with most recent estimates converging toward $d$$\approx$3~kpc
\citep[see][Figure~6]{hay10}, 
but its true value remains still rather uncertain.
The derived $^{44}$Ti mass as a function of the distance $d$ is shown in Fig.~4.
The red hatched area shows the region allowed by the BAT measurements. 
The blue hatched area includes the constraints from INTEGRAL \citep{renaud2}.
 
The double degenerate scenario, which invokes the dynamical merger of two WDs, can produce
$^{44}$Ti masses between 2\ee{-4} and 5\ee{-4} \msun, consistent with our values.
 This scenario would be disfavored by the presence of a donor star for SN~1572 \citep{ruiz04},
which, however, was recently questioned by \citet{nodonor}.

In Figure~4 we also report the predicted isotope masses for different 
types of explosion models in the single degenerate scenario.
Over the range of allowed distances, our results are broadly 
consistent with a sub-Chandrasekhar explosion producing a $^{44}$Ti yield between
\e{-4} and \e{-3} \msun. Extreme models, 
such as the double detonation (models 1-4; \citealt{fink10})
or the helium deflagration scenarios \citep{wk11}, are disfavored by the data. 

Typical Ia explosions, such as the fast deflagration W7 \citep{nomoto84}
or centrally-ignited pure deflagrations \citep{travaglio04}, 
predict masses $<$\e{-5}\,\msun, much lower than our inferred value.
For $d$$\lesssim$3.5 kpc, the BAT detection is consistent with some delayed detonation
explosions. In particular, the  WDD2, WDD3, and CDD2 models of \citet{iwamoto99},
which also predict the largest amount of $^{56}$Ni mass ($\sim$0.7\,\msun), and the
extremely off-center model O-DDT of \citet{maeda10}.
More recent three-dimensional simulations \citep{sss13} predicts lower nucleosynthetic yields,
marginally consistent with the derived range for $d \lesssim$2~kpc.
A delayed detonation explosion appears also consistent with the properties of the thermal 
X-ray emission \citep{b06}, and the SN 1572 light echo spectroscopy \citep{krause08}.

\begin{figure}
\centering
\includegraphics[scale=0.36]{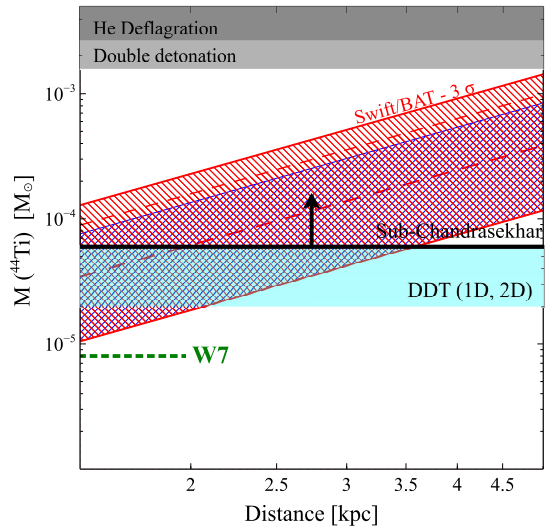}
\caption{Mass of the radio-isotope $^{44}$Ti, derived from the estimated line flux, as 
a function of the SNR distance.
The red hatched area shows the 1 $\sigma$ (dashed line), and 3$\sigma$ (solid line) 
confidence intervals from the {\it Swift}/BAT observations. The blue hatched area 
includes the 3\,$\sigma$ upper limit from
INTEGRAL observations. The mass range predicted by different explosion models is also reported.}
\end{figure}

\section{Summary and Conclusion}

The discovery of gamma-ray line emission at 1.157~MeV from the $^{44}$Ti-$^{44}$Sc-$^{44}$Ca 
decay chain in Cas ~A \citep{cgro94} was a benchmark observation for  
the field of astronomy with radioactivities 
\citep[see][for a recent review]{dhp07}, 
as it established the production and ejection of $^{44}$Ti from core collapse 
supernovae (ccSNe). 
The only other evidence for $^{44}$Ti production came from the late time light curve of SN ~1987A in the LMC, 
where modeling of the various contributions to power from radioactive decays indicated a $^{44}$Ti mass of 
02-2.0~\ee{-4}~\msun\ \citep[e.g.][]{fransson02}. The recent analysis of INTEGRAL observations of SN~1987A 
\citep{grebenev12} confirmed the optical/NIR lightcurve estimates, but supporting values for the Ti yield 
at the upper end of their range.

With a galaxy wide star formation rate of order 3~\msun~per year (see \citealt{diehl06} for a 
review and a discussion of the use of another radioactive tracer, $^{26}$Al, for measuring this rate) 
and a corresponding (IMF dependent) rate of 2-3 ccSNe per century one would expect several 
$^{44}$Ti hot spots produced during a few mean lives of this isotope. Monte Carlo simulations of the 
distribution of supernovae in the Milky Way \citep{the06}  indicated that the absence of additional $^{44}$Ti 
sources is indeed statistically unlikely. 
\citet{the06} concluded that either the Milky Way has experienced an unusually low core-collapse SN rate in the 
past few centuries, or that $^{44}$Ti producing supernovae are rare, atypical events.

The BAT detection of $^{44}$Ti in Tycho's supernova (Type Ia) would represent the third astrophysical site for which $^{44}$Ti synthesis 
(and ejection into the ISM) has been established, indicating that both core collapse supernovae and thermonuclear 
(single degenerate) supernovae contribute to the galactic chemical evolution of $^{44}$Ca. 
We now have two identified $^{44}$Ti sources in the galaxy (a ccSN remnant, Cas A, and a Type Ia remnant, Tycho) 
and one ccSN source in the LMC (SN 1987A), but this has not resolved the open question of the main source of $^{44}$Ca 
in galactic chemical evolution. The yields of the three sources discussed here are surprisingly similar, 
higher than the predictions of standard SN models, and at a level that led to the expected multi-spot sky 
simulated by \citet{the06}. 
We are therefore still missing the key to understanding the origin of $^{44}$Ca.




\bibliographystyle{aa}

\begin{thebibliography}{46}
\expandafter\ifx\csname natexlab\endcsname\relax\def\natexlab#1{#1}\fi

\bibitem[{{Ahmad} {et~al.}(2006){Ahmad}, {Greene}, {Moore}, {Ghelberg}, {Ofan},
  {Paul}, \& {Kutschera}}]{ahmad06}
{Ahmad}, I., {Greene}, J.~P., {Moore}, E.~F., {Ghelberg}, S., {Ofan}, A.,
  {Paul}, M., \& {Kutschera}, W. 2006, \prc, 74, 065803

\bibitem[{{Baade}(1945)}]{baade45}
{Baade}, W. 1945, \apj, 102, 309

\bibitem[{{Badenes} {et~al.}(2006){Badenes}, {Borkowski}, {Hughes}, {Hwang}, \&
  {Bravo}}]{b06}
{Badenes}, C., {Borkowski}, K.~J., {Hughes}, J.~P., {Hwang}, U., \& {Bravo}, E.
  2006, \apj, 645, 1373

\bibitem[{{Barthelmy} {et~al.}(2005){Barthelmy}, {Barbier}, {Cummings},
  {Fenimore}, {Gehrels}, {Hullinger}, {Krimm}, {Markwardt}, {Palmer},
  {Parsons}, {Sato}, {Suzuki}, {Takahashi}, {Tashiro}, \& {Tueller}}]{scott05}
{Barthelmy}, S.~D., {et~al.} 2005, \ssr, 120, 143

\bibitem[{{Baumgartner} {et~al.}(2013){Baumgartner}, {Tueller}, {Markwardt},
  {Skinner}, {Barthelmy}, {Mushotzky}, {Evans}, \& {Gehrels}}]{bat70}
{Baumgartner}, W.~H., {Tueller}, J., {Markwardt}, C.~B., {Skinner}, G.~K.,
  {Barthelmy}, S., {Mushotzky}, R.~F., {Evans}, P.~A., \& {Gehrels}, N. 2013,
  \apjs, 207, 19

\bibitem[{{Chieffi} \& {Limongi}(2013)}]{chieffi13}
{Chieffi}, A. \& {Limongi}, M. 2013, \apj, 764, 21

\bibitem[{{Clayton} {et~al.}(1969){Clayton}, {Colgate}, \&
  {Fishman}}]{clayton69}
{Clayton}, D.~D., {Colgate}, S.~A., \& {Fishman}, G.~J. 1969, \apj, 155, 75

\bibitem[{{Cusumano} {et~al.}(2010){Cusumano}, {La Parola}, {Segreto},
  {Mangano}, {Ferrigno}, {Maselli}, {Romano}, {Mineo}, {Sbarufatti}, {Campana},
  {Chincarini}, {Giommi}, {Masetti}, {Moretti}, \& {Tagliaferri}}]{palermo10}
{Cusumano}, G., {et~al.} 2010, \aap, 510, A48

\bibitem[{{Diehl} {et~al.}(2006){Diehl}, {Halloin}, {Kretschmer}, {Lichti},
  {Sch{\"o}nfelder}, {Strong}, {von Kienlin}, {Wang}, {Jean}, {Kn{\"o}dlseder},
  {Roques}, {Weidenspointner}, {Schanne}, {Hartmann}, {Winkler}, \&
  {Wunderer}}]{diehl06}
{Diehl}, R., {et~al.} 2006, \nat, 439, 45

\bibitem[{{Diehl} {et~al.}(2007){Diehl}, {Hartmann}, \& {Prantzos}}]{dhp07}
{Diehl}, R., {Hartmann}, D.~H., \& {Prantzos}, N. 2007, Meteoritics and
  Planetary Science, 42, 1145

\bibitem[{{Dupraz} {et~al.}(1997){Dupraz}, {Bloemen}, {Bennett}, {Diehl},
  {Hermsen}, {Iyudin}, {Ryan}, \& {Schoenfelder}}]{dupraz97}
{Dupraz}, C., {Bloemen}, H., {Bennett}, K., {Diehl}, R., {Hermsen}, W.,
  {Iyudin}, A.~F., {Ryan}, J., \& {Schoenfelder}, V. 1997, \aap, 324, 683

\bibitem[{{Eriksen} {et~al.}(2011){Eriksen}, {Hughes}, {Badenes}, {Fesen},
  {Ghavamian}, {Moffett}, {Plucinksy}, {Rakowski}, {Reynoso}, \&
  {Slane}}]{chandra}
{Eriksen}, K.~A., {et~al.} 2011, \apjl, 728, L28

\bibitem[{{Fink} {et~al.}(2010){Fink}, {R{\"o}pke}, {Hillebrandt},
  {Seitenzahl}, {Sim}, \& {Kromer}}]{fink10}
{Fink}, M., {R{\"o}pke}, F.~K., {Hillebrandt}, W., {Seitenzahl}, I.~R., {Sim},
  S.~A., \& {Kromer}, M. 2010, \aap, 514, A53

\bibitem[{{Fransson} \& {Kozma}(2002)}]{fransson02}
{Fransson}, C. \& {Kozma}, C. 2002, NewAR, 46, 487

\bibitem[{{Gehrels} {et~al.}(2004){Gehrels}, {Chincarini}, {Giommi}, {Mason},
  {Nousek}, {Wells}, {White}, {Barthelmy}, {Burrows}, {Cominsky}, {Hurley},
  {Marshall}, {M{\'e}sz{\'a}ros}, {Roming}, {Angelini}, {Barbier}, {Belloni},
  {Campana}, {Caraveo}, {Chester}, {Citterio}, {Cline}, {Cropper}, {Cummings},
  {Dean}, {Feigelson}, {Fenimore}, {Frail}, {Fruchter}, {Garmire}, {Gendreau},
  {Ghisellini}, {Greiner}, {Hill}, {Hunsberger}, {Krimm}, {Kulkarni}, {Kumar},
  {Lebrun}, {Lloyd-Ronning}, {Markwardt}, {Mattson}, {Mushotzky}, {Norris},
  {Osborne}, {Paczynski}, {Palmer}, {Park}, {Parsons}, {Paul}, {Rees},
  {Reynolds}, {Rhoads}, {Sasseen}, {Schaefer}, {Short}, {Smale}, {Smith},
  {Stella}, {Tagliaferri}, {Takahashi}, {Tashiro}, {Townsley}, {Tueller},
  {Turner}, {Vietri}, {Voges}, {Ward}, {Willingale}, {Zerbi}, \&
  {Zhang}}]{neil04}
{Gehrels}, N., {et~al.} 2004, \apj, 611, 1005

\bibitem[{{Grebenev} {et~al.}(2012){Grebenev}, {Lutovinov}, {Tsygankov}, \&
  {Winkler}}]{grebenev12}
{Grebenev}, S.~A., {Lutovinov}, A.~A., {Tsygankov}, S.~S., \& {Winkler}, C.
  2012, \nat, 490, 373

\bibitem[{{Grefenstette} {et~al.}(2014){Grefenstette}, {Harrison}, {Boggs},
  {Reynolds}, {Fryer}, {Madsen}, {Wik}, {Zoglauer}, {Ellinger}, {Alexander},
  {An}, {Barret}, {Christensen}, {Craig}, {Forster}, {Giommi}, {Hailey},
  {Hornstrup}, {Kaspi}, {Kitaguchi}, {Koglin}, {Mao}, {Miyasaka}, {Mori},
  {Perri}, {Pivovaroff}, {Puccetti}, {Rana}, {Stern}, {Westergaard}, \&
  {Zhang}}]{nustar14}
{Grefenstette}, B.~W., {et~al.} 2014, \nat, 506, 339

\bibitem[{{Harrison} {et~al.}(2013){Harrison}, {Craig}, {Christensen},
  {Hailey}, {Zhang}, {Boggs}, {Stern}, {Cook}, {Forster}, {Giommi},
  {Grefenstette}, {Kim}, {Kitaguchi}, {Koglin}, {Madsen}, {Mao}, {Miyasaka},
  {Mori}, {Perri}, {Pivovaroff}, {Puccetti}, {Rana}, {Westergaard}, {Willis},
  {Zoglauer}, {An}, {Bachetti}, {Barri{\`e}re}, {Bellm}, {Bhalerao},
  {Brejnholt}, {Fuerst}, {Liebe}, {Markwardt}, {Nynka}, {Vogel}, {Walton},
  {Wik}, {Alexander}, {Cominsky}, {Hornschemeier}, {Hornstrup}, {Kaspi},
  {Madejski}, {Matt}, {Molendi}, {Smith}, {Tomsick}, {Ajello}, {Ballantyne},
  {Balokovi{\'c}}, {Barret}, {Bauer}, {Blandford}, {Brandt}, {Brenneman},
  {Chiang}, {Chakrabarty}, {Chenevez}, {Comastri}, {Dufour}, {Elvis}, {Fabian},
  {Farrah}, {Fryer}, {Gotthelf}, {Grindlay}, {Helfand}, {Krivonos}, {Meier},
  {Miller}, {Natalucci}, {Ogle}, {Ofek}, {Ptak}, {Reynolds}, {Rigby},
  {Tagliaferri}, {Thorsett}, {Treister}, \& {Urry}}]{harrison13}
{Harrison}, F.~A., {et~al.} 2013, \apj, 770, 103

\bibitem[{{Hayato} {et~al.}(2010){Hayato}, {Yamaguchi}, {Tamagawa}, {Katsuda},
  {Hwang}, {Hughes}, {Ozawa}, {Bamba}, {Kinugasa}, {Terada}, {Furuzawa},
  {Kunieda}, \& {Makishima}}]{hay10}
{Hayato}, A., {et~al.} 2010, \apj, 725, 894

\bibitem[{{Hwang} {et~al.}(2002){Hwang}, {Decourchelle}, {Holt}, \&
  {Petre}}]{hwang02}
{Hwang}, U., {Decourchelle}, A., {Holt}, S.~S., \& {Petre}, R. 2002, \apj, 581,
  1101

\bibitem[{{Iwamoto} {et~al.}(1999){Iwamoto}, {Brachwitz}, {Nomoto},
  {Kishimoto}, {Umeda}, {Hix}, \& {Thielemann}}]{iwamoto99}
{Iwamoto}, K., {Brachwitz}, F., {Nomoto}, K., {Kishimoto}, N., {Umeda}, H.,
  {Hix}, W.~R., \& {Thielemann}, F.-K. 1999, \apjs, 125, 439

\bibitem[{{Iyudin} {et~al.}(1994){Iyudin}, {Diehl}, {Bloemen}, {Hermsen},
  {Lichti}, {Morris}, {Ryan}, {Schoenfelder}, {Steinle}, {Varendorff}, {de
  Vries}, \& {Winkler}}]{cgro94}
{Iyudin}, A.~F., {et~al.} 1994, \aap, 284, L1

\bibitem[{{Kerzendorf} {et~al.}(2013){Kerzendorf}, {Yong}, {Schmidt}, {Simon},
  {Jeffery}, {Anderson}, {Podsiadlowski}, {Gal-Yam}, {Silverman}, {Filippenko},
  {Nomoto}, {Murphy}, {Bessell}, {Venn}, \& {Foley}}]{nodonor}
{Kerzendorf}, W.~E., {et~al.} 2013, \apj, 774, 99

\bibitem[{{Kothes} {et~al.}(2006){Kothes}, {Fedotov}, {Foster}, \&
  {Uyan{\i}ker}}]{kothes06}
{Kothes}, R., {Fedotov}, K., {Foster}, T.~J., \& {Uyan{\i}ker}, B. 2006, \aap,
  457, 1081

\bibitem[{{Krause} {et~al.}(2008{\natexlab{a}}){Krause}, {Birkmann}, {Usuda},
  {Hattori}, {Goto}, {Rieke}, \& {Misselt}}]{krause08}
{Krause}, O., {Birkmann}, S.~M., {Usuda}, T., {Hattori}, T., {Goto}, M.,
  {Rieke}, G.~H., \& {Misselt}, K.~A. 2008{\natexlab{a}}, Science, 320, 1195

\bibitem[{{Krause} {et~al.}(2008{\natexlab{b}}){Krause}, {Tanaka}, {Usuda},
  {Hattori}, {Goto}, {Birkmann}, \& {Nomoto}}]{krausetycho}
{Krause}, O., {Tanaka}, M., {Usuda}, T., {Hattori}, T., {Goto}, M., {Birkmann},
  S., \& {Nomoto}, K. 2008{\natexlab{b}}, \nat, 456, 617

\bibitem[{{Lampton} {et~al.}(1976){Lampton}, {Margon}, \& {Bowyer}}]{lampton76}
{Lampton}, M., {Margon}, B., \& {Bowyer}, S. 1976, \apj, 208, 177

\bibitem[{{Leising}(2001)}]{leising01}
{Leising}, M.~D. 2001, \apj, 563, 185

\bibitem[{{Maeda} {et~al.}(2010){Maeda}, {R{\"o}pke}, {Fink}, {Hillebrandt},
  {Travaglio}, \& {Thielemann}}]{maeda10}
{Maeda}, K., {R{\"o}pke}, F.~K., {Fink}, M., {Hillebrandt}, W., {Travaglio},
  C., \& {Thielemann}, F.-K. 2010, \apj, 712, 624

\bibitem[{{Nomoto} {et~al.}(1984){Nomoto}, {Thielemann}, \& {Yokoi}}]{nomoto84}
{Nomoto}, K., {Thielemann}, F.-K., \& {Yokoi}, K. 1984, \apj, 286, 644

\bibitem[{{Renaud} {et~al.}(2006{\natexlab{a}}){Renaud}, {Vink},
  {Decourchelle}, {Lebrun}, {den Hartog}, {Terrier}, {Couvreur},
  {Kn{\"o}dlseder}, {Martin}, {Prantzos}, {Bykov}, \& {Bloemen}}]{renaud06}
{Renaud}, M., {et~al.} 2006{\natexlab{a}}, \apjl, 647, L41

\bibitem[{{Renaud} {et~al.}(2006{\natexlab{b}}){Renaud}, {Vink},
  {Decourchelle}, {Lebrun}, {Terrier}, \& {Ballet}}]{renaud2}
{Renaud}, M., {Vink}, J., {Decourchelle}, A., {Lebrun}, F., {Terrier}, R., \&
  {Ballet}, J. 2006{\natexlab{b}}, NewAR, 50, 540

\bibitem[{{Reynolds} \& {Keohane}(1999)}]{rk99}
{Reynolds}, S.~P. \& {Keohane}, J.~W. 1999, \apj, 525, 368

\bibitem[{{Ruiz-Lapuente} {et~al.}(2004){Ruiz-Lapuente}, {Comeron},
  {M{\'e}ndez}, {Canal}, {Smartt}, {Filippenko}, {Kurucz}, {Chornock}, {Foley},
  {Stanishev}, \& {Ibata}}]{ruiz04}
{Ruiz-Lapuente}, P., {et~al.} 2004, \nat, 431, 1069

\bibitem[{{Segreto} {et~al.}(2010){Segreto}, {Cusumano}, {Ferrigno}, {La
  Parola}, {Mangano}, {Mineo}, \& {Romano}}]{segreto10}
{Segreto}, A., {Cusumano}, G., {Ferrigno}, C., {La Parola}, V., {Mangano}, V.,
  {Mineo}, T., \& {Romano}, P. 2010, \aap, 510, A47

\bibitem[{{Seitenzahl} {et~al.}(2013){Seitenzahl}, {Ciaraldi-Schoolmann},
  {R{\"o}pke}, {Fink}, {Hillebrandt}, {Kromer}, {Pakmor}, {Ruiter}, {Sim}, \&
  {Taubenberger}}]{sss13}
{Seitenzahl}, I.~R., {et~al.} 2013, \mnras, 429, 1156

\bibitem[{{Slane} {et~al.}(2014){Slane}, {Lee}, {Ellison}, {Patnaude},
  {Hughes}, {Eriksen}, {Castro}, \& {Nagataki}}]{slane14}
{Slane}, P., {Lee}, S.-H., {Ellison}, D.~C., {Patnaude}, D.~J., {Hughes},
  J.~P., {Eriksen}, K.~A., {Castro}, D., \& {Nagataki}, S. 2014, \apj, 783, 33

\bibitem[{{Tamagawa} {et~al.}(2009){Tamagawa}, {Hayato}, {Nakamura}, {Terada},
  {Bamba}, {Hiraga}, {Hughes}, {Hwang}, {Kataoka}, {Kinugasa}, {Kunieda},
  {Tanaka}, {Tsunemi}, {Ueno}, {Holt}, {Kokubun}, {Miyata}, {Szymkowiak},
  {Takahashi}, {Tamura}, {Ueno}, \& {Makishima}}]{suzaku}
{Tamagawa}, T., {et~al.} 2009, \pasj, 61, 167

\bibitem[{{The} {et~al.}(2006){The}, {Clayton}, {Diehl}, {Hartmann}, {Iyudin},
  {Leising}, {Meyer}, {Motizuki}, \& {Sch{\"o}nfelder}}]{the06}
{The}, L.-S., {et~al.} 2006, \aap, 450, 1037

\bibitem[{{Travaglio} {et~al.}(2004){Travaglio}, {Hillebrandt}, {Reinecke}, \&
  {Thielemann}}]{travaglio04}
{Travaglio}, C., {Hillebrandt}, W., {Reinecke}, M., \& {Thielemann}, F.-K.
  2004, \aap, 425, 1029

\bibitem[{{Tueller} {et~al.}(2010){Tueller}, {Baumgartner}, {Markwardt},
  {Skinner}, {Mushotzky}, {Ajello}, {Barthelmy}, {Beardmore}, {Brandt},
  {Burrows}, {Chincarini}, {Campana}, {Cummings}, {Cusumano}, {Evans},
  {Fenimore}, {Gehrels}, {Godet}, {Grupe}, {Holland}, {Kennea}, {Krimm},
  {Koss}, {Moretti}, {Mukai}, {Osborne}, {Okajima}, {Pagani}, {Page}, {Palmer},
  {Parsons}, {Schneider}, {Sakamoto}, {Sambruna}, {Sato}, {Stamatikos},
  {Stroh}, {Ukwata}, \& {Winter}}]{bat22}
{Tueller}, J., {et~al.} 2010, \apjs, 186, 378

\bibitem[{{Vink}(2012)}]{vink2012}
{Vink}, J. 2012, \aapr, 20, 49

\bibitem[{{Vink} {et~al.}(2001){Vink}, {Laming}, {Kaastra}, {Bleeker},
  {Bloemen}, \& {Oberlack}}]{vink2001}
{Vink}, J., {Laming}, J.~M., {Kaastra}, J.~S., {Bleeker}, J.~A.~M., {Bloemen},
  H., \& {Oberlack}, U. 2001, \apjl, 560, L79

\bibitem[{{Wang} \& {Li}(2014)}]{wangli14}
{Wang}, W. \& {Li}, Z. 2014, \apj, 789, 123

\bibitem[{{Woosley} \& {Kasen}(2011)}]{wk11}
{Woosley}, S.~E. \& {Kasen}, D. 2011, \apj, 734, 38

\bibitem[{{Woosley} \& {Weaver}(1994)}]{ww94}
{Woosley}, S.~E. \& {Weaver}, T.~A. 1994, \apj, 423, 371

\end{thebibliography}

\end{document}